\documentclass[aps,prl,twocolumn,showpacs,superscriptaddress,groupedaddress]{revtex4} 
\usepackage{graphicx}  
\usepackage{dcolumn}   
\usepackage{bm}        
\usepackage{amssymb}   

\usepackage[latin1]{inputenc}
\usepackage{tikz}
\usetikzlibrary{shapes,arrows}

\usepackage{graphicx,amsfonts,epsfig,latexsym,amscd,amsmath,theorem,mathrsfs,color}
\usepackage{subcaption}
\usepackage{ragged2e}

\def\comment#1{}

\def\cm#1{}

\newcommand{\hh}{{\cal H}}
\newcommand{\eps}{\epsilon}

\newcommand{\be}{\begin{equation}}
\newcommand{\ee}{\end{equation}}
\newcommand{\ba}{\begin{eqnarray}}
\newcommand{\ea}{\end{eqnarray}}
\newcommand{\beq}{\begin{equation}}
\newcommand{\eeq}{\end{equation}}
\newcommand{\bea}{\begin{eqnarray}}
\newcommand{\eea}{\end{eqnarray}}
\newcommand{\bastar}{\begin{eqnarray*}}
\newcommand{\eastar}{\end{eqnarray*}}

\newcommand{\cd}{\partial}

\newcommand{\ignore}[1]{}

\begin{document}

\widetext

\title{Glass transitions in monodisperse cluster-forming ensembles:\\ vortex matter in type-1.5 superconductors}

\author{Rogelio D\'iaz-M\'endez}%
\affiliation{icFRC, IPCMS (UMR 7504), ISIS (UMR 7006), Universit\'e de Strasbourg and CNRS, 67000 Strasbourg, France}%
\author{Fabio Mezzacapo}%
\affiliation{icFRC, IPCMS (UMR 7504), ISIS (UMR 7006), Universit\'e de Strasbourg and CNRS, 67000 Strasbourg, France}%
\author{Wolfgang Lechner}%
\affiliation{IQOQI and Institute for Theoretical Physics, University of Innsbruck, Austria}%
\author{Fabio Cinti}%
\affiliation{National Institute for Theoretical Physics (NITheP), Stellenbosch 7600, South Africa}%
\affiliation{Institute of Theoretical Physics, Stellenbosch University, Stellenbosch 7600, South Africa}
\author{Egor Babaev}%
\affiliation{Department of Theoretical Physics and Center for Quantum Materials, KTH-Royal Institute of Technology,
Stockholm, SE-10691 Sweden}
\author{Guido Pupillo}%
\affiliation{icFRC, IPCMS (UMR 7504), ISIS (UMR 7006), Universit\'e de Strasbourg and CNRS, 67000 Strasbourg, France}%
 
\date{\today}

\begin{abstract}
At low enough temperatures and high densities, the equilibrium configuration of an ensemble of ultrasoft particles is a
self-assembled, ordered, cluster-crystal. 
In the present work we explore the out-of-equilibrium dynamics for a
two-dimensional realisation, which is relevant to  superconducting materials with multi-scale 
intervortex forces. We find that for small 
temperatures following a quench, the suppression of the thermally-activated particle hopping hinders the ordering. This results in a
glass transition for a {\it monodispersed} ensemble, for which we derive a microscopic explanation in terms
of an ``effective polydispersity" induced by multi-scale interactions. This demonstrates that a vortex glass can form in clean systems of thin films of   ``type-1.5" superconductors. An additional setup to study this physics can be layered superconducting systems, where the shape of the effective vortex-vortex interactions can be engineered. 
\end{abstract}

\maketitle
\begin{figure}[t]
\centerline{\includegraphics[height=.7\columnwidth,angle=-0]{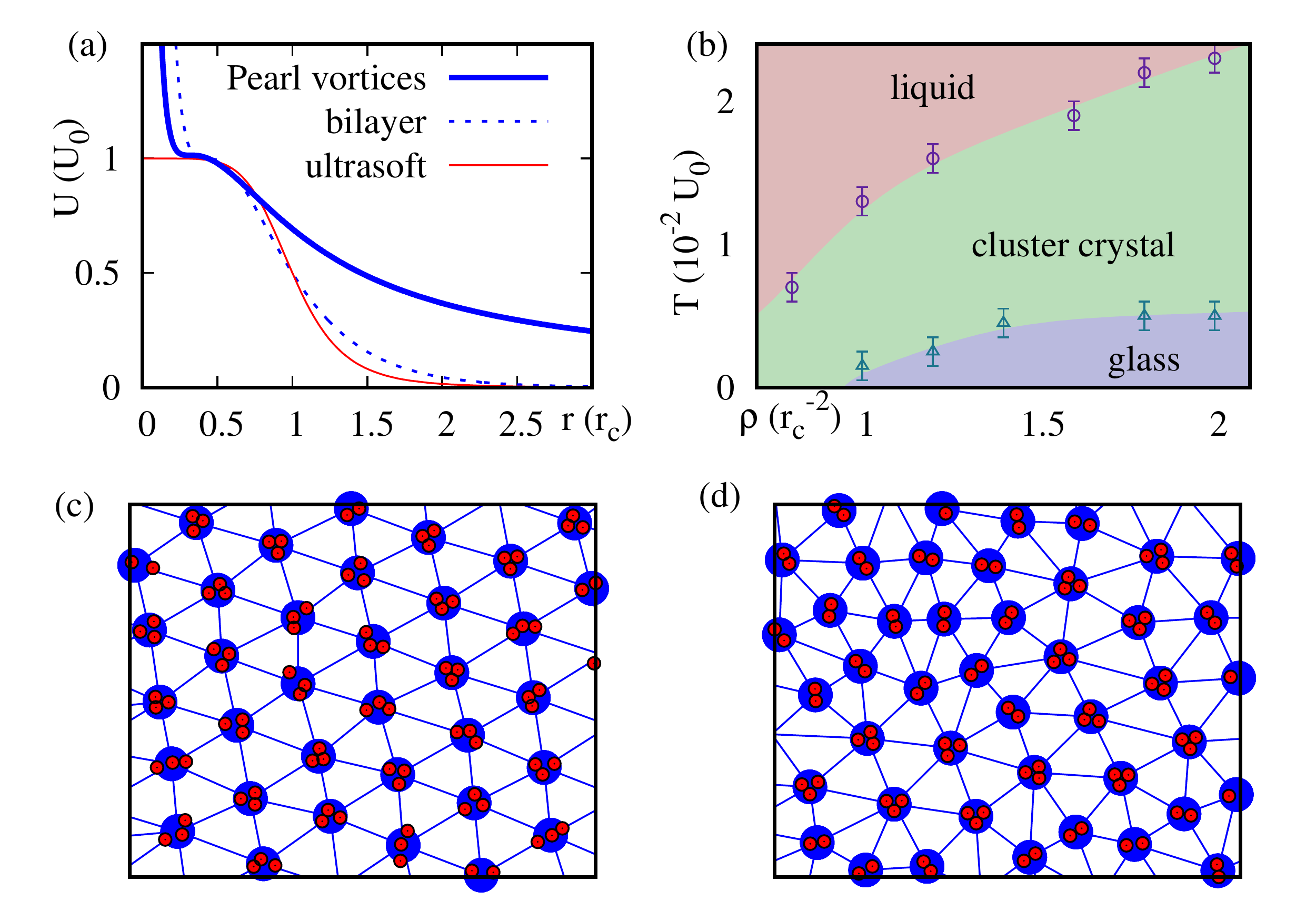}}
\caption{
a) Cluster-forming interaction for a system of vortices in 1.5 superconducting films [thick blue line, Eq. ~(\ref{eintfilm})
in the text]. 
It is also shown the interaction potential for vortices in 1.5 bilayer superconductors [dashed blue line,
Eq.~(\ref{eq:vortex0}) in the text], and the generic ultrasoft potential [red line, Eq.~(\ref{u}) in the text]. 
b)~Dynamic phase diagram of the model of Eq.~(\ref{eintfilm}) as a function of rescaled
density and temperature.  Circles and triangles indicate the liquid-to-crystal and the glass transition
temperatures, respectively.
c)~Snapshot 
of a crystal configuration after quenching a monodisperse vortex system with the potential 
Eq.(\ref{eintfilm}) [blue line in panel (a)], for density $r_c^2\rho=1.6$ at temperature $T=1.8\times10^{-2}U_0$.
Single vortices (red circles) group into clusters (blue circles), blue lines join nearest neighboring clusters as obtained by Delauney triangulation.
d)~Same as (c) for the glass phase at density $r_c^2\rho=1.6$ and temperature $T=0.2\times10^{-2}U_0$. In all simulations we choose the values $A/U_0 = 0.7364$, 
$C_B^2/U_0= 8.124$, 
$\lambda/r_c= 0.0084$, 
$C_1^2/U_0= 0.884$ and
$\xi_1/r_c= 0.238$.  
}
\label{f1}
\end{figure}

The vortex glass is one of the key states in the theory of  magnetic and transport properties of type-2 superconductors in the presence of disorder~\cite{fisher1989vortex,fisher1991thermal,blatter1994vortices,nelson2002defects}.
Within this frame, the glassy phase
is caused  by the pinning of vortices by impurities and is absent in a clean sample.
In this work we demonstrate that a vortex glass state can be an inherent property of a superconducting system 
characterised by multiple coherence lengths.
In a more general context, our work demonstrates that a structurally disordered glass state of matter can be 
obtained in the absence of disordered substrates for a simple two-dimensional monodisperse ensemble 
of particles interacting via isotropic, repulsive, ultrasoft interactions \cite{malescio07,likos06}. 
This is surprising as those conditions are usually associated with minimal frustration \cite{berthier10,ikeda11,ikeda11b,coslovich2012}. 
The glass phase appears below a non-equilibrium glass transition temperature  and extends to the lowest temperatures examined. 
We provide a description of the microscopic mechanism responsible for the appearance of glassiness in terms of an {\it effective polydispersity} that emerges following a quench due to multi-scale interactions [see Fig.~\ref{f1}].
In typical glass forming liquids, frustration results from polydisperse mixtures of particles \cite{binder2011}. 
In the present systems, the appearance of glassiness stems from the effective polydispersity of clusters sizes. 

Several  works have recently discussed ``type-1.5 superconductors'' that are characterised by multiple coherence lengths, some of which are larger and some smaller than the magnetic field penetration length. 
These multiple coherence lengths arise in superconducting states that break multiple symmetries and also in materials with multiple superconducting bands. 
Several materials were suggested in experiments to belong to this type of superconductors \cite{moshchalkov,gutierrez2012scanning,Ray.Gibbs.ea:14,noncentr}, where
vortices can display multi-scale attractive and repulsive inter-vortex interactions~\cite{bs11,Babaev.Carlstrom.ea:10,silaev1,moshchalkov,Ray.Gibbs.ea:14}.
Multiple attractive length scales come from core-core intervortex interactions. 
Multiple repulsive length scales can be obtained instead in (i) artificially fabricated superconducting bilayers,  where the different layers give rise to two coherence lengths, or rather generally in (ii) thin films of type-1.5 materials due to  stray fields.

In the case of artificial superconducting bilayers [case (i) above], the London's magnetic field penetration length will in general be different in the different layers. In this case, co-centered vortices form in different layers in the presence of a perpendicular magnetic field.
When the interlayer electromagnetic coupling is strong (or there is substantial interlayer proximity effect),
fluctuations associated with the loss of axial symmetry of vortices can, under certain conditions, be neglected.  
For a sufficiently high vortex line tension, a dilute system of such vortices can be mapped onto point particles with the following intervortex
interaction potential at long-ranges derived in the SupMat
\footnote{See Supplemental Material, which includes Refs. \cite{silaev2, tilley, gurevich, gurevich2, zhitomirsky, garaud2016microscopically, frac, npb, sublattice, smiseth2005field, chung2010entropy, bs1, johan1, johan2, spe, carneiro2000vortex,varney13,Silaev.Babaev:11}.}
\begin{equation}
 U(r)=\sum_{i=1,2} \left[C_{B_i}^2 K_0\left(\frac{r}{\lambda_i}\right)
     -C_i^2  K_0\left(\frac{r}{\xi_i}\right)\right].
\label{eq:vortex0}
\end{equation}
Here, $K_0$ is the modified Bessel function of the second kind. 
In SupMat we show simulations for such a system 
with different London's magnetic field penetration lengths $\lambda_{1,2}$, coherence lengths $\xi_{1,2}$, and coefficients
$C_{B_i}$ and $C_i$, which are weakly dependent on $T$ (see SupMat).
The resulting potential shape for that particular choice is shown in Fig.~\ref{f1}(a) with a blue dashed line.

In the case of   films of type-1.5 superconductors  [case (ii) above]
the long range interaction potential acquires a term that decays with distance as $1/r$ due to  Pearl's effect~\cite{pearl},
\beq
U(r)= {C_{B}^2} K_0\left(\frac{r}{{\lambda}}\right) + \frac{A}{r}-\sum_i{C_i^2} K_0\left(\frac{r}{\xi_i}\right)
\label{eintfilm}
\eeq
Here, like in the ordinary films of type-2 superconductors we separated the electromagnetic interaction in the interior of the film and the Pearl's $A/r$ correction arising from demagnetisation fields (A being a constant). 
The prefactors $C_{B}$ and $C_i$, in Eq.~(\ref{eintfilm}) depend on the film thickness, while ${\lambda}$ and $\xi_i$ depend on the choice of material.  
Both potentials in Eqs.~(\ref{eq:vortex0}) and~(\ref{eintfilm}) [thin dashed and thick blue curves in Fig.~\ref{f1}(a), respectively] 
are two-scale repulsive, in contrast to the single-scale repulsive inter-vortex potentials in usual type-2 superconductors.
In their range of validity they have  a ``plateau''-like feature of value $U_0$ at intermediate distances that extends up to a distance
$r\simeq r_c$, and then decays for $r \gtrsim r_c$. Note that  for type-1.5 vortices one can have
a more pronounced minimum in the place of the  plateau which yields similar results. Here we are interested in the dynamics of two-dimensional vortices following a  quench from an initial high-temperature to final low-temperature $T$, for vortex densities $\rho$ such as $r_c^2\rho\gtrsim 1$. We perform molecular dynamics simulations with an overdamped Langevin thermostat of friction coefficient $\gamma$, governed by the equation $ \dot{\vec{r}}=-\nabla U/\gamma+\sqrt{2k_BT/\gamma}\ \vec{\eta}(t)$, where $\vec{\eta}(t)$ is a gaussian stochastic force with zero mean and unit variance and $k_B$ is the Boltzmann constant~\cite{Tinkham2004, Silaev2013, Bardeen1965}.
The units of length, time, temperature and density are $r_c$, $\gamma^{-1}$,  $U_0$ and $r_c^{-2}$, respectively, and the
total number of particles $N$ varies from $N=600$ to 15000. 
\footnote{For the larger systems we used the GPU-acceleration provided by the HOOMD-Blue simulation toolkit \cite{hoomd1,
hoomd2}.}

Figure~\ref{f1}(b) shows the phase diagram of Eq.~(\ref{eintfilm}) following the temperature quench to a final value $T$ [see below for details on observables].
For comparatively large $T$ the system remains in a liquid phase, while for intermediate temperatures it equilibrates and the resulting phase
is a cluster-crystal [see snapshot in panel (c)].  In this phase particles group into clusters, which in turn are ordered in a triangular lattice with approximately the same number of particles per site. 
For the lowest temperatures, we find instead a surprising lack of equilibration that keeps the system in a disordered configuration [see panel (d)]. The demonstration of a resulting vortex glass in the absence of substrate disorder and its microscopic explanation in terms of effective polydispersity (see below) are main results of this work. 

 We note that, by a further increase of the density with $\rho r_c^2\gg 1$, the low-$T$ configuration in both models above can evolve through  states 
 induced due to multiple interaction scales.
At equilibrium these do not resemble simple triangular cluster crystals. 
As an example, the formation of a disordered nematic-like phase for the bilayer model of
Eq.~(\ref{eq:vortex0}) and density $r_c^2\rho=4.0$ is shown in SupMat. 

Since the divergence of the 
interaction potential at $r=0$
is an artefact of asymptotical analysis, in the following we also consider a model of cluster-forming potential where the unphysical short-range divergence is removed. 
We come back to the potentials of Eqs.~(\ref{eq:vortex0}) and~(\ref{eintfilm}) in SupMat. 
This model potential reads
\begin{equation}
U(r)=U_0 \left[1+\left(r/r_c\right)^6\right]^{-1}.
\label{u}
\end{equation}
Such a potential approaches the constant value $U_0$ as the inter-particle distance $r$ decreases below the soft-core
radius $r_c$, and drops to zero for $r>r_c$ as $r^{-6}$ i.e., with a repulsive van der Waals tail
\cite{maucher2011}.
Ultrasoft potentials of this kind have recently attracted considerable attention~\cite{mladek2006, coslovich2011,
montes2013, sciortino2013} as mean-field approximations of inter-polymer interactions in soft-matter systems as diverse
as dendritic polymers, polymer rings, and chains. 
Due to their negative Fourier components~\cite{likos2001,likos2007, glaser2007}, they  provide an unexpected route towards self-assembly of composite crystalline structures for sufficiently large densities. 

\begin{figure}[t]
\centerline{\includegraphics[width=1\columnwidth]{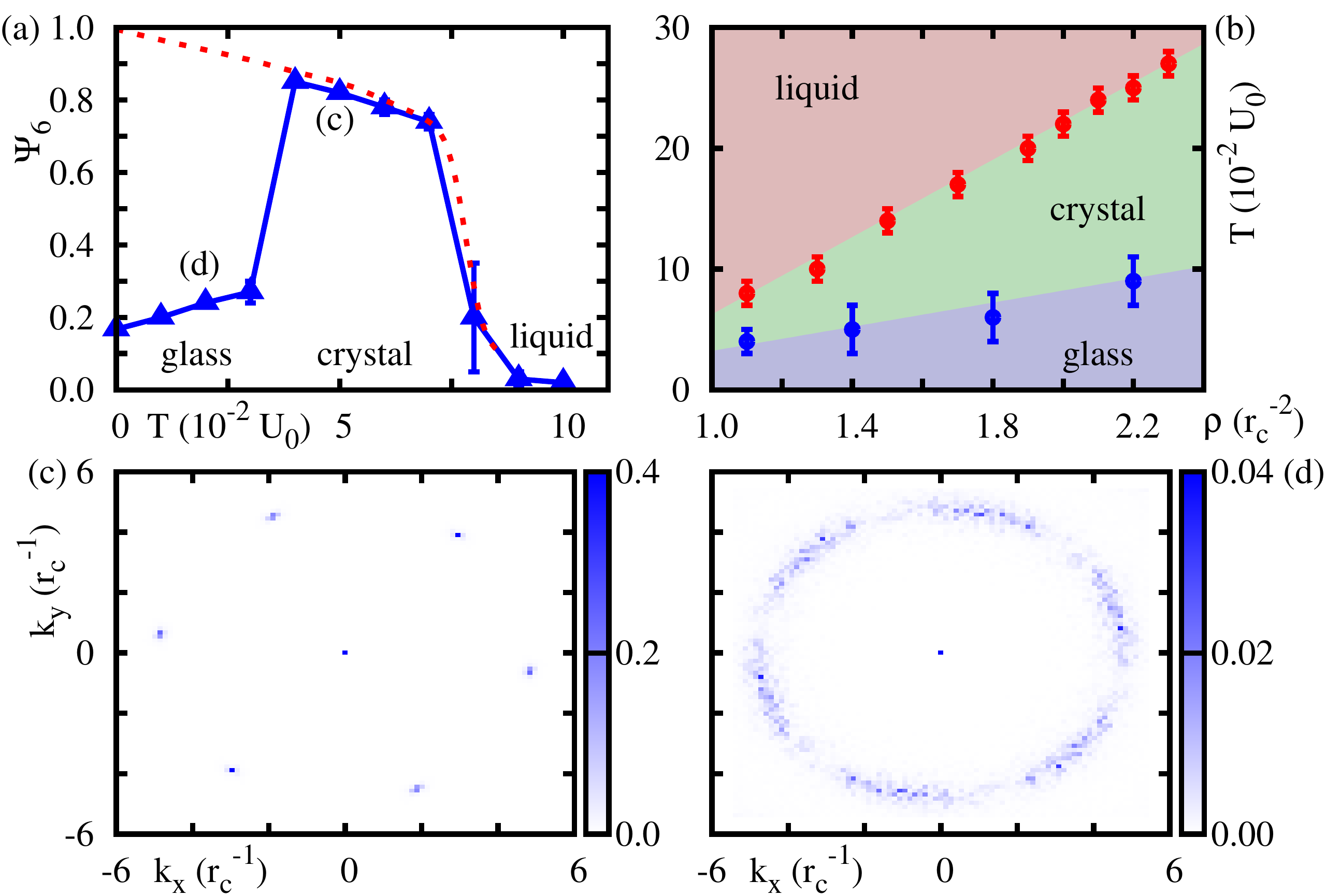}}
\caption{
a) Hexatic order parameter $\psi_6$ as a function of the temperature for a system of particles interacting via the potential
in Eq.~(\ref{u}) at equilibrium, and after a quench from high $T$ (red dashed line and blue triangles, respectively). 
The chosen value of the particle density is $r_c^2\rho=1.1$. b) Dynamic phase diagram of the model as a function of rescaled
density and temperature.  Red and blue symbols indicate the liquid-to-crystal and the glass transition temperatures, respectively. 
Typical structure factors of the system are shown for the crystalline [panel (c)] and disordered glass [panel (d)] phase.
}
\label{f2}
\end{figure}

The phase diagrams for all models following a temperature quench are determined by computing both static and dynamical observables, corresponding to the hexatic order parameter for clusters  $\Psi_6=\langle\sum_j^{N_c}\sum_l^{N_j}e^{i6\theta_{jl}}/(N_cN_j)\rangle$ (see Ref.~\cite{ourselves}), the static structure factor $S(\mathbf k)=\langle|\sum_{j}^N e^{i\mathbf{k}\cdot\mathbf{r}_j}|^2/N\rangle $, the mean-square displacement $\langle\Delta r^2(t)\rangle=\langle \sum_j|\mathbf{r}_j(0)-\mathbf{r}_j(t)|^2\rangle/N$ and the non-gaussian parameter  $\alpha_2(t)=\left[\langle\Delta r^4(t)\rangle/(2\langle\Delta r^2(t)\rangle^2)-1\right]$. 
Here, angular brackets $\langle \cdot \rangle$ denote an average over quench experiments, $N_c$ is the total number of clusters, $N_j$ is the coordination number of cluster $j$ (corresponding to the number of clusters neighboring the $j$-th one),  $\theta_{jl}$ is the angle between a reference axis and the segment joining the clusters $j$ and $l$ (see SupMat and Ref.~\cite{ourselves}), and $t$ is time.


Figure~\ref{f2} shows results for the ultrasoft model of Eq.~(\ref{u}).
Panel (a) shows the values of the hexatic order parameter $\Psi_6$ at equilibrium (red dashed line, see
Ref.~\cite{ourselves}), and after a quench (blue triangles) as a function of the final temperature $T$ of the system, for a
fixed density. 
The equilibrium results display a single sudden jump of $\Psi_6$ from 0 to about 0.8 at $T_c \simeq 8\times10^{-2}U_0$, followed by a slow rise to 1 with decreasing $T$.
This jump corresponds to a transition from the high-temperature disordered liquid to an ordered cluster-crystalline phase for $T<T_c$. Each cluster here comprises the same time-averaged number of particles.
The finite value of $S(\mathbf k)$ in this finite system
reflects the quasi-long range 
order of the crystal. In contrast, the results following the
temperature quench display two jumps. The first, at  $T_c \simeq 8\times10^{-2}U_0$, corresponds to the onset of crystal
formation: For $T_g < T < T_c$ 
equilibration into a regular cluster crystal occurs in a time scale 
much smaller than the simulation time and the crystal
is essentially indistinguishable from the equilibrium situation [see Fig.~\ref{f2}(a) and Fig.~\ref{f2}(c)].
Conversely, the second 
jump is 
characteristic of the quenched dynamics, and corresponds to the onset of glass formation: Below a characteristic temperature
$T_g$ the value of $\Psi_6$ suddenly becomes small, signalling disorder.  Disappearance of structural order is further
demonstrated by
the formation of a ring-like feature in $S(\bf{k})$ [Fig.~\ref{f2}(d)]. 
By inspection (see below), we find that disorder here results from the loss of ergodicity and the consequent lack of equilibration within the
simulation time: Following the quench, particles quickly re-arrange in clusters, however particle hopping between clusters is suppressed, so that the distribution of particles among the clusters remains disperse in time. This suggests that inter-cluster interactions, which depend on cluster occupancies, can vary significantly in the ensemble. As a result, clusters do not evolve into a large isotropic crystalline structure below $T_g$.

\begin{figure}[t]
\includegraphics[width=\columnwidth]{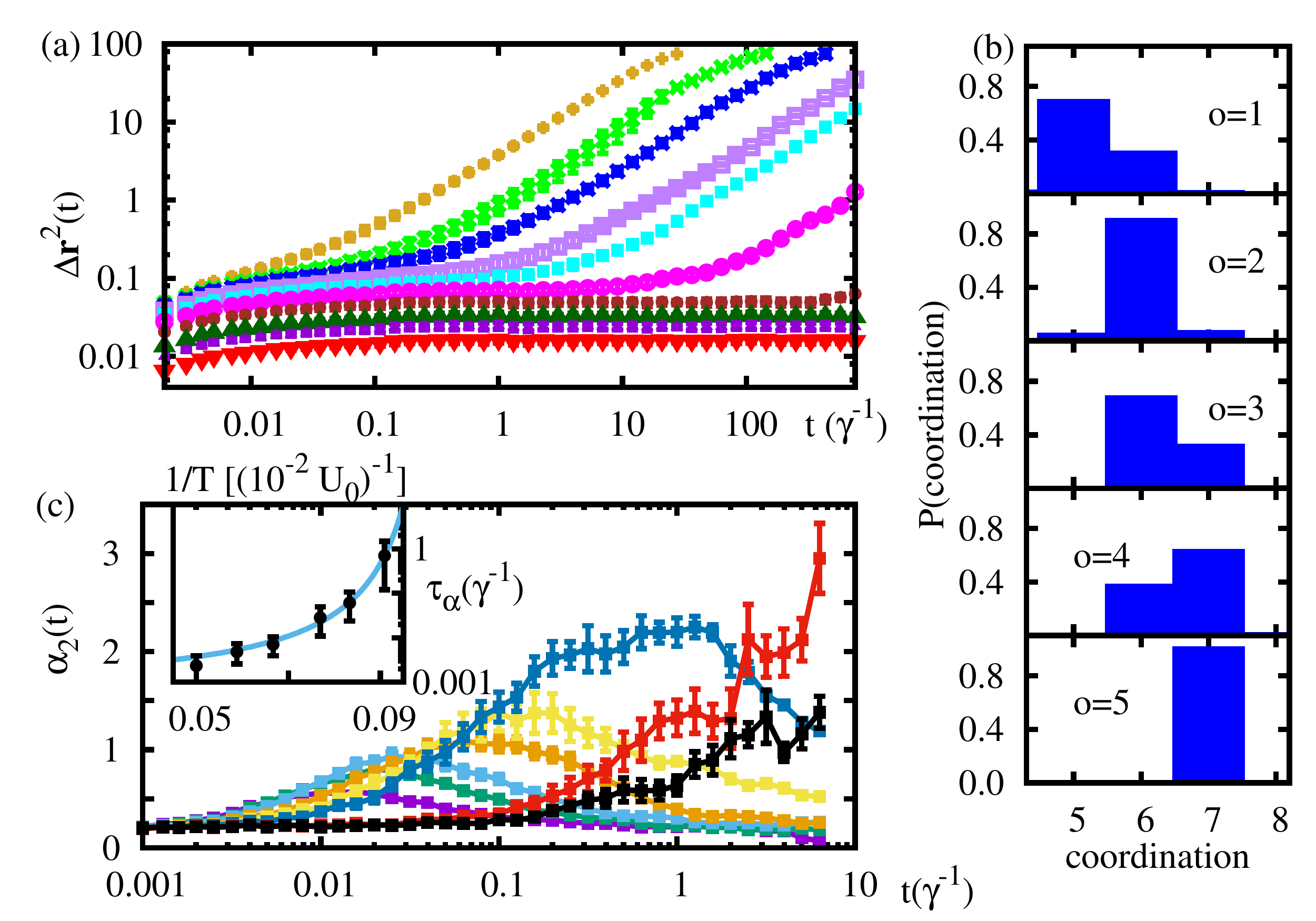}
\caption{
a) Mean square displacement for a system of particles interacting via Eq.~(1) with $r_c^2\rho=1.1$, and temperatures, from
top to bottom, $T=\{10, 9, 8, ..., 1\}\times10^{-2} U_0$. 
b) Probability distribution of the coordination number (i.e., the
number of  clusters nearest neighboring a given one)  for subsets of clusters having the same occupation number $o$ ( i.e.,
the number of particles in the cluster), for a choice of parameters so that the system is a glass.
c) Non-gaussian  parameter $\alpha_2(t)$ for density $\rho=1.4$. 
The temperatures (from the most left to the most right position of the maximum) are $T=\{12, 11, 9, 8, 7, 6, 5,
3\}\times10^{-2}U_0$. 
The inset shows the Arrhenius plot of the relaxation time, as extracted from the maximum of the $\alpha_2(t)$ parameter. 
}
\label{f3}
\end{figure}

The emerging picture is one where an \emph{effective polydispersity} of the clusters is realised in this low-temperature
regime, corresponding to the formation of clusters with different occupancies. 
Our interpretation of the microscopic mechanism of induced polydispersity is quantified in the data of Fig.~\ref{f3}(b). The
latter presents histograms of the measured coordination
numbers of the clusters (i.e., the number of clusters which are nearest neighbor of a given cluster) for different cluster
occupation values (i.e., number of particles in a cluster). 
The figure shows that the smaller the cluster occupation, the higher is the probability for a given cluster of being
low-coordinated, and vice versa. In other words, small ``less repulsive'' clusters are more likely to have less
neighbors than large ``more repulsive'' ones. This strongly suggests a correlation between the induced ``effective polydispersity" and the
structural disorder of the glassy phase. 
An analogy can be drown here with the effect of particle size distribution in the formation of
disordered structures  in genuinely
polydisperse ensembles \cite{weeks2000,tanaka2010,markland2011,Ebert2008,binder2011,Lechner2013}. 
Interestingly, the equilibrium counterpart of this glass is  a crystal, which turns into a glass for quenches at target
temperature lower than  $T_g$. We note that glass transitions have been previously found as a
function of the degree of polidispersity in certain quasi-two-dimensional samples of binary colloidal suspensions
\cite{yunker2010}.  The development of glassy properties in those polydisperse models  is in some respects similar to the
behaviour found in our monodispersed ensemble, though here glassiness originates solely from the multiscale interactions.

We confirm the glassy dynamics by monitoring time-dependent observables such as the  mean square displacement  $\langle\Delta r^2(t)\rangle$ and  $\alpha_2(t)$.
Example results for
$r_c^2 \rho >1$ are shown in Fig.~3(a). In the figure, a linear dependence of $\langle\Delta r^2(t)\rangle$ on time $t$ signals liquid
behaviour, which is evident for high $T$ (yellow line). For intermediate temperatures, however,  $\langle\Delta r^2(t)\rangle$ develops a
plateau at intermediate times. The latter is usually associated to caging effects when, close to a glass transition, mobility
of individual particles is increasingly limited. In our case, this behaviour occurs in the intermediate
temperature range where the system rearranges in a cluster crystalline configuration. Here the long-time liquid-like dynamics
corresponds to residual activated particle hopping between the cluster sites, as observed first in
Ref.~\cite{ourselves}.
Interestingly, for lower temperatures the dynamics after a quench is completely arrested (i.e.,  $\langle\Delta r^2(t)\rangle$ takes a low value essentially constant in $t$), 
consistent with 
a transition to a glassy phase. 

In Fig.~3(c) we plot the non-gaussian parameter $\alpha_2(t)$ as a function of $t$ and for several values of $T$.
This parameter measures deviations from gaussian fluctuations in the distribution of displacements, and thus is in
general $\alpha_2(t)=0$ for all $t$ in regular liquids and non-cluster crystals at equilibrium. Here, at intermediate temperatures 
$\alpha_2(t)$ takes a maximum for a characteristic time $t=\tau_\alpha$.  
This signals the presence of different time scales
usually associated with dynamical heterogeneity and out-of-equilibrium glassy dynamics. 
Our estimates for $\tau_\alpha$ are consistent with a
Vogel-Fulcher-Tamman dependence on temperature
[inset in Fig.~3(c)],  which, within the structural glass-forming liquid scenario, usually indicates a fragile
nature of the glass transition. 
The glass phase is found to extend down to the lowest temperatures probed. 
In SupMat we show that the energy barriers that prohibit hopping between clusters are heterogeneous, which is a typical signature of glasses \cite{binder2011}.

We find that the freezing temperature 
derived from the time dependent quantities
$\langle\Delta
r^2(t)\rangle$ and $\alpha_2(t)$  is in agreement with 
the glass transition temperature $T_g$
obtained
from the static observables $\Psi_6$ and $S(\mathbf k)$. 
This should make the experimental observation of the glass phase possible directly from snapshots of
particle distributions. 
The demonstration of a glass phase in a low dimensional monodisperse system with purely repulsive and isotropic
inter-particle interactions in free space and its explanation in terms of induced polydispersity is one of the central results of this work.\\

In summary, we have demonstrated the existence of glass transitions in monodisperse isotropic systems without  disorder. 
The mechanism for this unusual glass formation has been identified as a consequence of multi-scale interaction potentials. 
While geometrical frustration in typical glass forming materials stems from polydispersity of particles, here, disorder is an
effective consequence of frustration in the hopping in the context of cluster-crystals and therefore a distribution of
various cluster sizes. 
The transition is a two-step process: first, the clusters form, and then, in a second step, they order. 
It is this second step that shows glassy dynamics due to the effective polydispersity of cluster sizes.
One of the physical consequences is that a vortex glass state of matter is possible in clean systems: namely in thin films of type-1.5 superconductors. It can also be realized  in artificial layered materials that can provide new
experimental venues  to explore soft-matter models with microscopic control of interactions. 

R. D. M., F. M. and G. P. acknowledge  support by the European Commission via ERC-St Grant ColdSIM
(No. 307688) and Rysq, UdS via Labex NIE and IdEX, computing time at the HPC-UdS. W. L. acknowledges support by the Austrian
Science Fund through Grant No. P 25454-N27 and by the Institut fuer Quanteninformation. 
E.B. acknowledges support from Goran Gustafsson Foundation  and by the Swedish
Research Council 642-2013-7837.

{\it Note:} after completion of this work, we became aware of the related work \cite{miyazaki16} on a cluster glass transition
in a model with binary mixtures in three dimensions.


\pagebreak
\widetext
\begin{center}
\textbf{\large Supplemental Material to: ``Glass Transitions in Monodisperse Cluster-Forming Ensembles: Vortex Matter in Type-1.5 Superconductors''}
\end{center}
\setcounter{equation}{0}
\setcounter{figure}{0}
\setcounter{table}{0}
\setcounter{page}{1}
\makeatletter
\renewcommand{\theequation}{S\arabic{equation}}
\renewcommand{\thefigure}{S\arabic{figure}}
\renewcommand{\bibnumfmt}[1]{[#1]}
\renewcommand{\citenumfont}[1]{#1}

\section{Inter-vortex interaction potential}
In this Supplemental Material, we derive the effective inter-vortex interaction potential for type-1.5 superconductors of Eq.~(1) of the main text. Calculations similar to the ones outlined below can also be 
carried in microscopic models \cite{silaev1,silaev2}.
Here we begin by  considering the following multi-component  Ginzburg-Landau (GL) functional that describes two superconducting components coupled by Josephson term
\begin{equation}
F=\sum_{i,j=1,2}\frac{1}{2}(D\psi_i)(D\psi_i)^* 
+\frac{1}{2}(\nabla\times {\bf A})^2 + \eta|\psi_i||\psi_j|\cos({\theta_i-\theta_j})+V_p.
\label{gl}
\end{equation}
Here,
$D=\nabla + ie {\bf A}$, 
$e$ is the  coupling constant that in these units parametrises the magnetic field penetration length
and $\psi_i=|\psi_a| e^{i\theta_i}$,
$i=1,2$, represent   superconducting  components either in different bands or for example in different proximity-effect-coupled layers.
The terms $\eta|\psi_i||\psi_j|\cos({\theta_i-\theta_j})$ represent interlayer or interband Josephson-like coupling.
The term $V_p$ contains potential terms.  In the simplest case it
has the form $V_p=\sum_i a_i|\psi_i|^2+\frac{b_i}{2}|\psi_i|^4$.
The  Eq.~(\ref{gl}) can be  obtained 
as an expansion in small gaps and small gradients from microscopic models \cite{tilley,gurevich,gurevich2,zhitomirsky,silaev2,garaud2016microscopically}. 
The only vortex solutions with finite energy per unit length are the ones with similar phase winding:
i.e. for integrals around the composite vortex core with $\oint_\sigma \nabla \theta_i= 2\pi N$ (see detailed discussion in \cite{frac}). Below a certain characteristic temperature 
one can neglect fluctuations associated with 
relative phase gradients or splitting of the vortex cores (see
estimates e.g. in \cite{npb,sublattice,smiseth2005field,chung2010entropy}).
Then, unless the superconductor is type-1, the minimal vortex energy per flux quantum corresponds to a vortex
that carries one flux quantum: i.e. $N=1$.\\

The long-range intervortex forces can be found 
by  linearization of the theory.
Following \cite{bs1,johan1,johan2} we consider the case where we have $2$-component superconductors 
with phase differences  locked to zero $\theta_1=\theta_2=0$
and the ground states are given by 
$|\psi_i|=u_i$. Consider a vortex  in the 2-band model.
Then we can write
\begin{align}
 \psi_i&=f_i(r)e^{i\theta}\,,&
(A_1,A_2)&=\frac{a(r)}{r}(-\sin\theta,\cos\theta)\label{ansatz}
\end{align}
where $f_i,a$ have the following behavior
$f_i(0)=a(0)=0$, $f_i(\infty)=u_i$, $a(\infty)=-1/e$. 
For studying long range inter-vortex forces,
one can linearise the model and  
derive inter-vortex interactions using the source method. 
The latter method is described  in detail for a single component GL theory in \cite{spe}.
The inter-vortex interaction at large separation coincides with
that between the corresponding point-like perturbations interacting via the linearized
field theory. For Eq.~(\ref{gl}), the linearization has one vector
(${\bf A}$) and three  real
scalar fields ($\eps_i=|\psi_i|-u_i$ and $\theta_i-\theta_j$).

Linearizing the GL equations for small $\eps_i$ we get 
\be
F_{lin}=\sum_i\frac12|\nabla\eps_i|^2 +
\frac12\left(\begin{array}{c}\eps_1\\ \eps_2 \end{array}\right)\cdot
\hh\left(\begin{array}{c}\eps_1\\ \eps_2\end{array}\right)
\label{gllin}
+\frac12(\cd_1A_2-\cd_2A_1)^2
+\frac12e^2u_i^2|A|^2.
\ee
Here,
\beq
\hh_{ij}=\left.\frac{\cd^2F_p}{\cd|\psi_i|\cd|\psi_j|}\right|_{(u_i)}.
\eeq
The vector potential decouples and
gives the  London's magnetic field penetration length  $\lambda$: 

\beq
\lambda^{-1}=e\sqrt{\sum_i u_i^2}.
\eeq

The attractive interaction  between vortices is due to
core overlaps.
For the superconducting component density fields
we can remove the cross-terms by a linear transformation.
As a result we obtain \be
F_{lin}=\frac12\sum_{i=1}^2\left(|\nabla\chi_i|^2+\mu_a^2\chi_a^2\right)
+\frac12(\cd_1A_2-\cd_2A_1)^2
+\frac12e u_i^2|A|^2.
\ee
The  fields $\chi_i,\chi_2$  are linear combinations 
of the original density fields that recover ground state values at 
the scale of 
coherence length scales $\xi_i =1/\mu_i$. 
From this theory, 
the form of the long-range inter-vortex interaction energy can be
determined  \cite{bs1,johan1,johan2,spe}.
Vortices here can be viewed as point particles associated with the 
centers of their cores.
 
The interaction energy of  two such vortices separated by a distance $r$
is 
\beq
E_{int}= {m^2} K_0(\mu_A r)-\sum_i{q_i^2} K_0(\mu_i r)
\label{eint1}
\eeq
where 
$K_0$ denotes the modified Bessel's function of the second kind. 
Here the first term describes the repulsive inter-vortex
interaction. The repulsive length scale given by the  magnetic field penetration length is $\lambda=\mu_A^{-1}$. 
The other two terms describe the attractive interaction. The ranges of the two coherence lengths are given by $\xi_i$. 
The coefficients $m$ and $q_i$ are determined by nonlinearities and can be found numerically from the full non-linear model.

\begin{figure}[t]
\centerline{\includegraphics[width=0.3 \columnwidth]{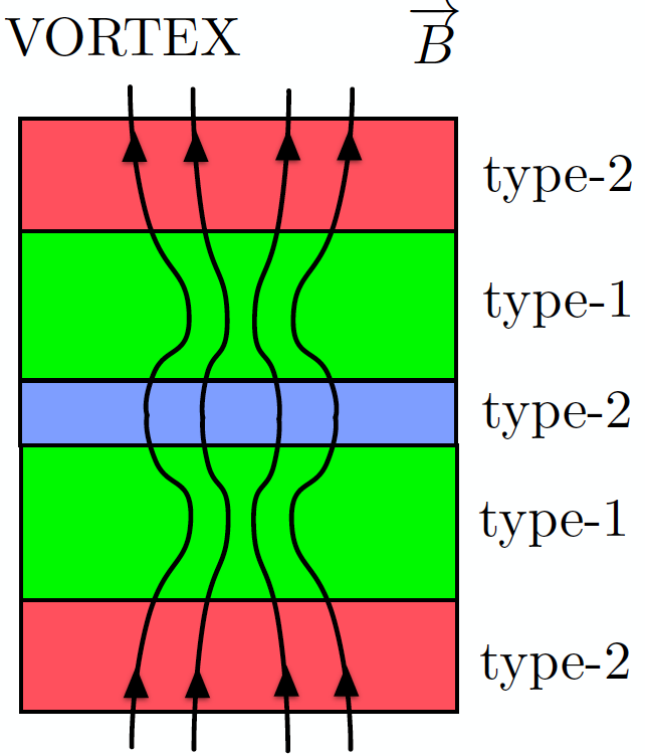}}
\caption{ A schematic picture of a multilayer made of superconductors with different
$\lambda$ and $\xi$. Different localization of magnetic field in different layers
gives multi-scale interaction for each vortex stack.
}
\label{fs}
\end{figure}

We note that since the equation above is obtained using asymptotical analysis, it becomes incorrect at the origin $r\to 0$, where this interaction energy
diverges. The actual inter-vortex potential would instead saturate to a finite value. 
While the exact inter-vortex interaction potential at all distances can be in principle obtained numerically, as explained in \cite{johan2},
we note that for our purposes the very short-range interaction is not important. In fact, a saturation of the interactions
at short distances would only enhance the clustering behaviour described in the main text.\\


The analysis above can be further generalized to the case of multiple repulsive length scales \cite{varney13}.
To this end, we consider  vortices in a layered system that are induced  by a magnetic 
field perpendicular to the layers as shown on Fig.~\ref{fs}.
For simplicity, we consider the case where each layer
is made from a superconductor with one coherence length and the layer thickness is $L_\alpha$.
In a multilayer system, vortices in different layers are coupled electromagnetically and in the ground state are straight lines \cite{blatter1994vortices}.
For fields perpendicular to the layers and at temperatures below those associated with pancake-vortex fluctuations~\cite{blatter1994vortices}, neglecting the contribution from interfaces between the layers the inter-vortex interaction becomes
\beq
E_{int}=L_\alpha{m^{(\alpha)}{}^2}K_0(\mu_A^{(\alpha)} r)-L_\alpha {q^{(\alpha)}{}^2} K_0(\mu^{(\alpha)} r).
\label{eint2}
\eeq

\beq
E_{int}=\sum_\alpha L_\alpha{m^{(\alpha)}{}^2}K_0(\mu_A^{(\alpha)} r)-L_\alpha{q^{(\alpha)}{}^2}K_0(\mu^{(\alpha)} r).
\label{eint3}
\eeq

There are many natural tuning parameters in these systems.
The layer thicknesses can be controlled in the fabrication process and 
the coefficients $\mu_A^{(\alpha)}$ and $\mu^{(\alpha)}$ by the choice of material.
This opens up a possibility to fabricate a desired intervortex 
interaction potential. 

%


Here we used Ginzburg-Landau model. As noted above, similar calculations of multi-scale inter-vortex 
interaction potentials can also be performed starting from a microscopic theory \cite{Silaev.Babaev:11}
giving the same form of interaction.


\subsection{Simulations for bilayers of type-1.5 superconductors}

In the main text the interaction potential of the multilayered system is presented in the form

\begin{equation}
 \small
 U(r)=\sum_{i=1,2} \left[C_{B_i}^2 K_0\left(\frac{r}{\lambda_i}\right)
     -C_i^2  K_0\left(\frac{r}{\xi_i}\right)\right].
\label{q:vortex0}
\end{equation}

For the simulations we focus on two layers with  $\lambda_1=0.42 r_c$ and $\lambda_2=0.14 r_c$. 
The latter set the range of the two scales of electromagnetic and current-current interaction 
in the two layers. 
The parameters $\xi_{1,2}$ are two coherence lengths that set the range of core-core interactions. 
Here, we consider the case where $\xi_2 \ll \xi_1$ so that we can neglect the contribution of the second core in the intervortex interaction, 
and choose $\xi_1=0.21 r_c$. 
The rest of the coefficients are set to $C_{B_1}=3.05 U_0^{1/2}$, $C_{B_2}=6.69 U_0^{1/2}$ and $C_1=6.26 U_0^{1/2}$. 
This particular potential is shown in the Fig.~1(a) of the main text with a dashed blue line. 

\begin{figure}[t]
\includegraphics[width=.6\columnwidth]{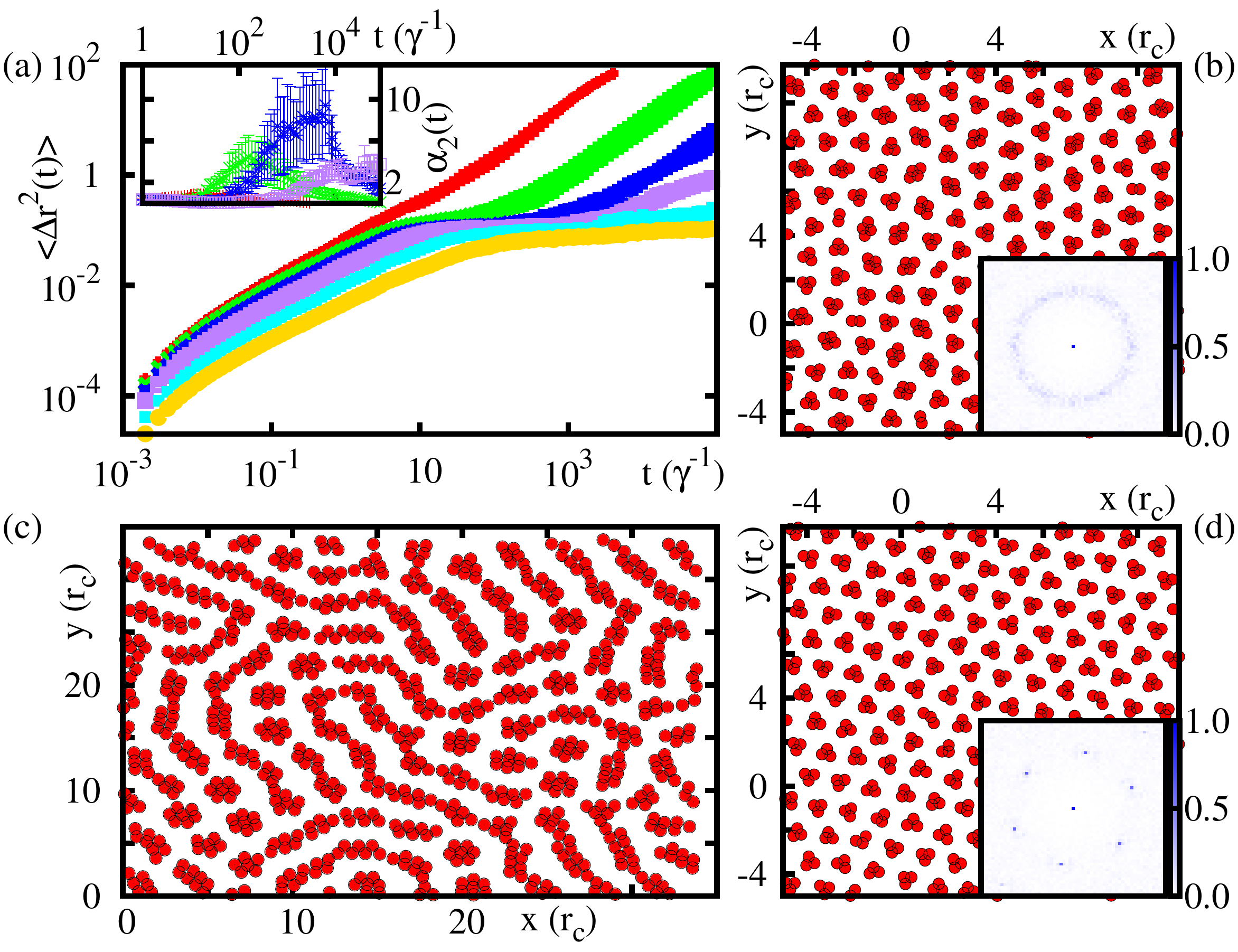}
\caption{a) Mean square displacement after quenches to temperatures, from top to bottom, $T=\{5, 4, 3, 2, 1, 0.5\}\times10^{-2}U_0$, for a system of vortices interacting via the potential in Eq. (\ref{q:vortex0}) at density $r_c^2\rho=1.8$.
The corresponding non-gaussian parameter is shown in the inset.
Panel (b) and (d) are typical snapshots and structure factors of the system in the glassy ($T=0.5\times10^{-2}U_0$) and  crystalline ($T=3\times10^{-2}U_0$) phases, respectively.
Panel (c) shows a typical configuration at high density, here $r_c^2\rho=4.0$. 
}
\label{f4}
\end{figure}

Figure~\ref{f4}(a) shows temperature-quench results for the mean square displacement $\langle\Delta r^2(t)\rangle$ and the
non-gaussian parameter $\alpha_2(t)$ as a function of time $t$ for such a vortex ensemble with density $r_c^2\rho = 1.8$.  
In complete analogy with the picture presented in the main text, at high and intermediate $T$ the system equilibrates and the resulting phases are a liquid characterised by the
linear dependence of the mean square displacement on time, and a cluster-crystal [see snapshot in panel (d)], where
$\langle\Delta r^2(t)\rangle$ displays an extended plateau at intermediate $t$ with linear diffusion recovered at large $t$.
For the lowest temperatures  shown, the lack of equilibration keeps the system in a disordered configuration [see snapshot in panel (b)]. 
In such a state the dynamics is arrested as shown by the mean square displacement, which takes a low value essentially constant in the limit of large $t$. 

We note that, by increasing the density further, the low-$T$ configuration of this superconductor model can further evolve through different self-assembled  states. 
As an example, panel (c) shows the formation of a disordered nematic-like phase of clusters  obtained at density $r_c^2\rho=4.0$. 
In our calculations the divergences of the vortex potentials were completely included, since for the range of temperatures and densities explored in this work the minimum distance between vortices corresponded to finite, numerically tractable forces. 
At larger densities the cluster crystal structure  is destabilized in favor of other self-assembled states whose detailed study is beyond the scope of the present work. 

\section{Energy barriers}

\begin{figure}
\begin{subfigure}[t]{.4\textwidth}
\centering
\includegraphics[width=\linewidth]{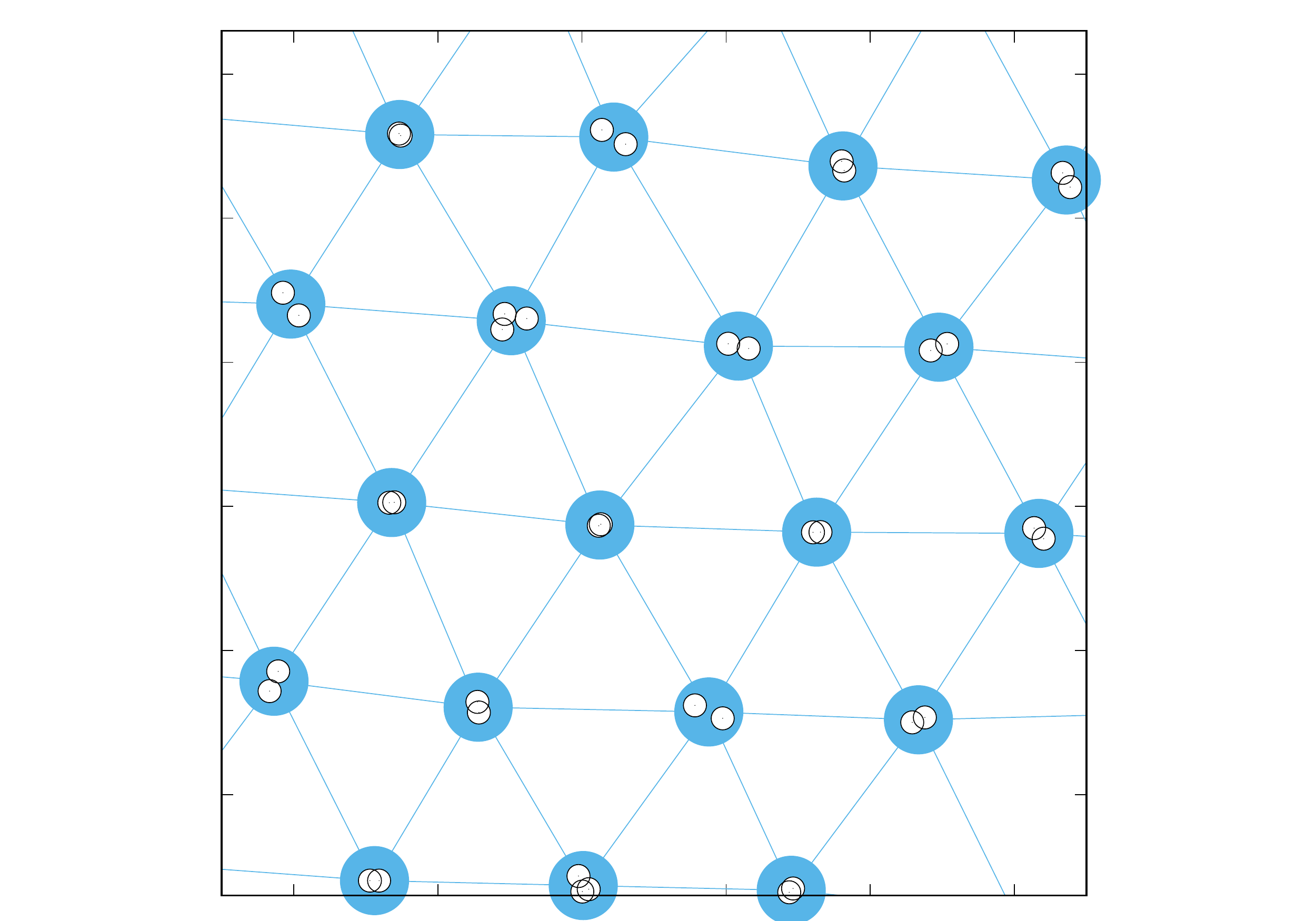}
        \caption{}\label{fig:fig_a}
\end{subfigure}
\begin{subfigure}[t]{.4\textwidth}
\centering
\includegraphics[width=\linewidth]{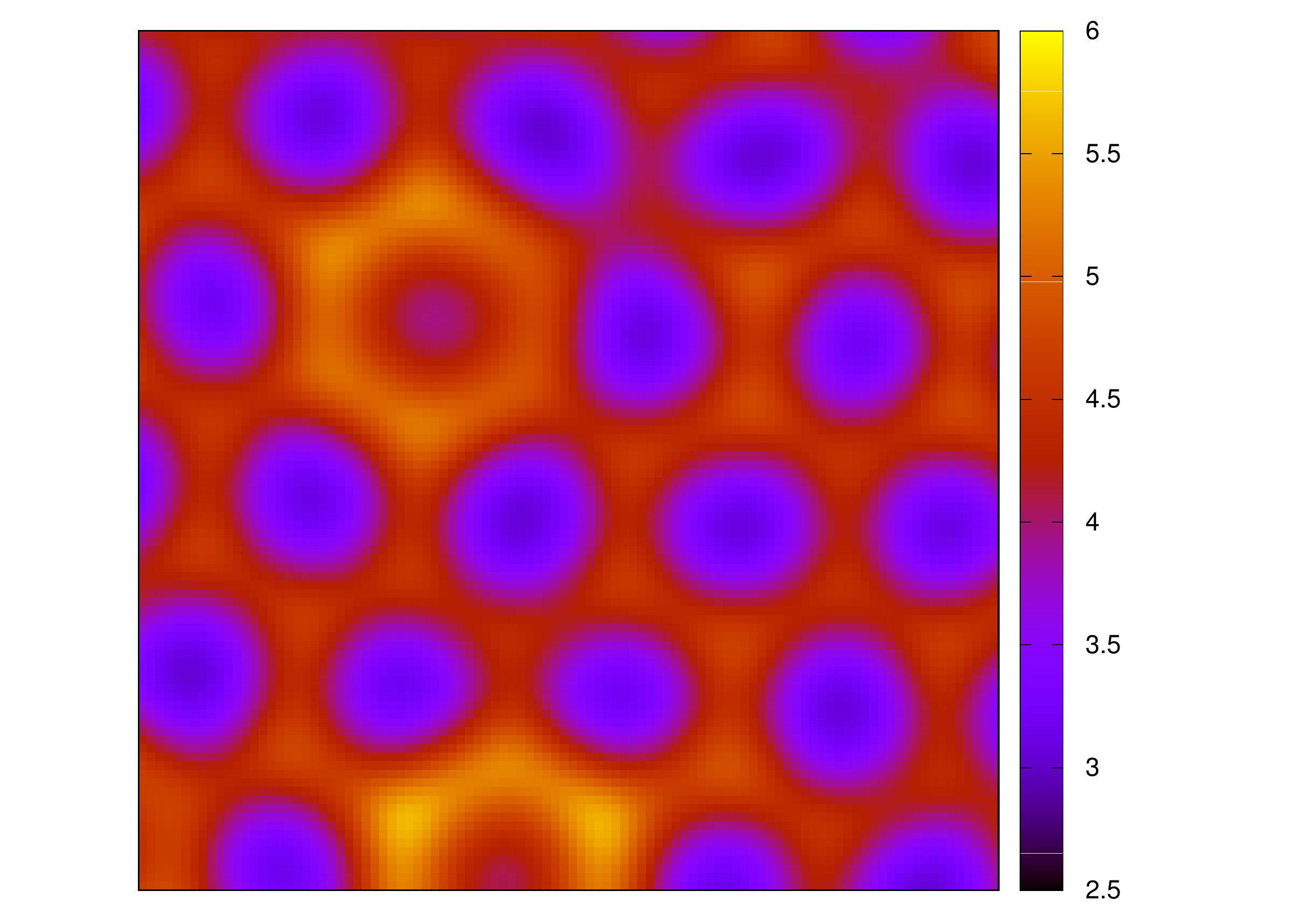}
\caption{}\label{fig:fig_b}
\end{subfigure}
\begin{subfigure}[t]{.4\textwidth}
\centering
\includegraphics[width=\linewidth]{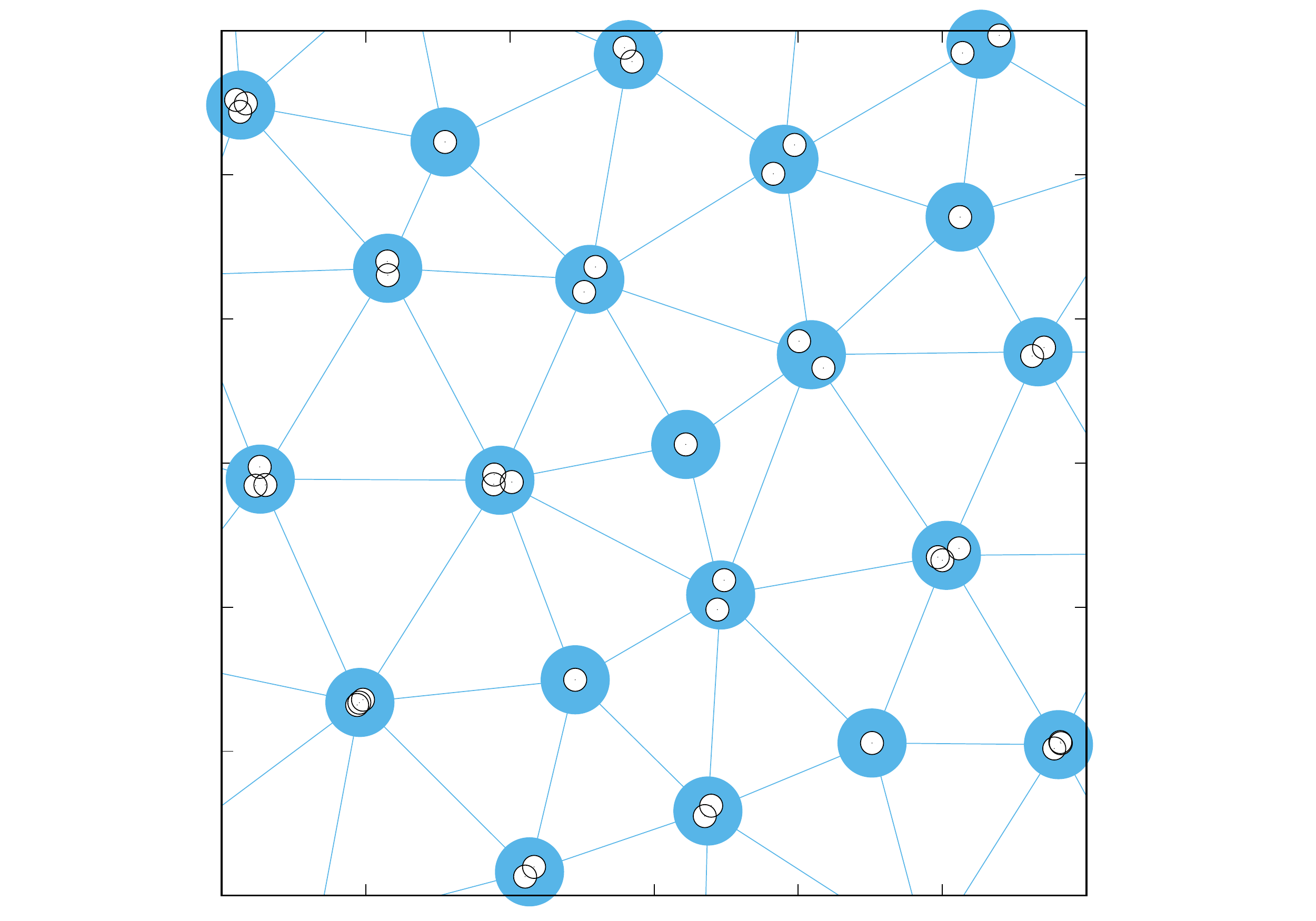}
\caption{}\label{fig:fig_a}
\end{subfigure}
\begin{subfigure}[t]{.4\textwidth}
\centering
\includegraphics[width=\linewidth]{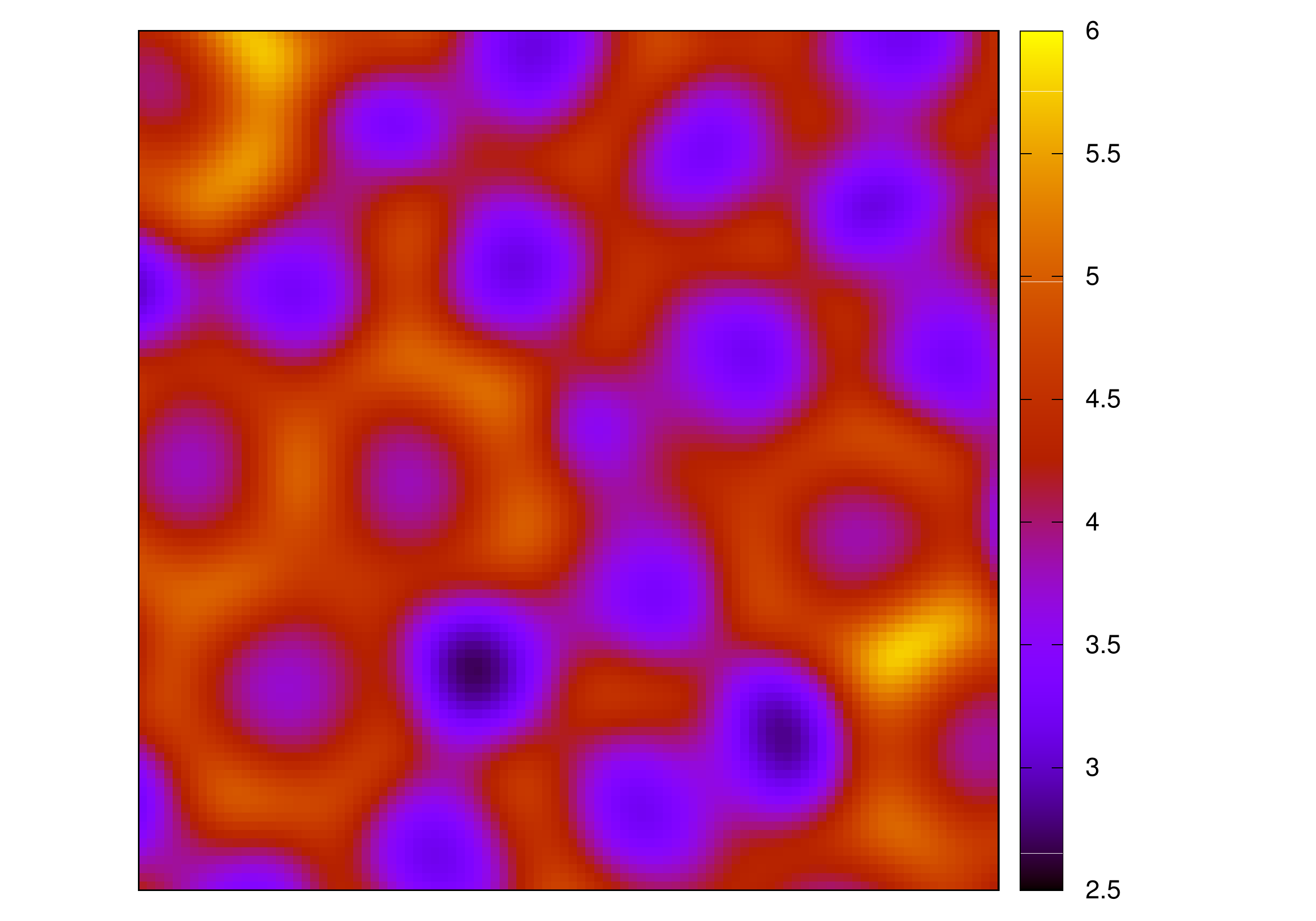}
\caption{}\label{fig:fig_b}
\end{subfigure}
\caption{Ultrasoft model in the crystal [a) and b)] and glass [c) and d)] phases. 
a) and c) are snapshots with clusters and nearest-neighbors connections remarked in cyan.
b) and d) are the corresponding maps of potential energy. 
In the disordered phase, clusters are connected via a heterogenoeous distribution of energy barriers, which is a typical signature for the glass phase. 
}
\label{ultra}
\end{figure}

\begin{figure}
\begin{subfigure}[t]{.4\textwidth}
\centering
\includegraphics[width=\linewidth]{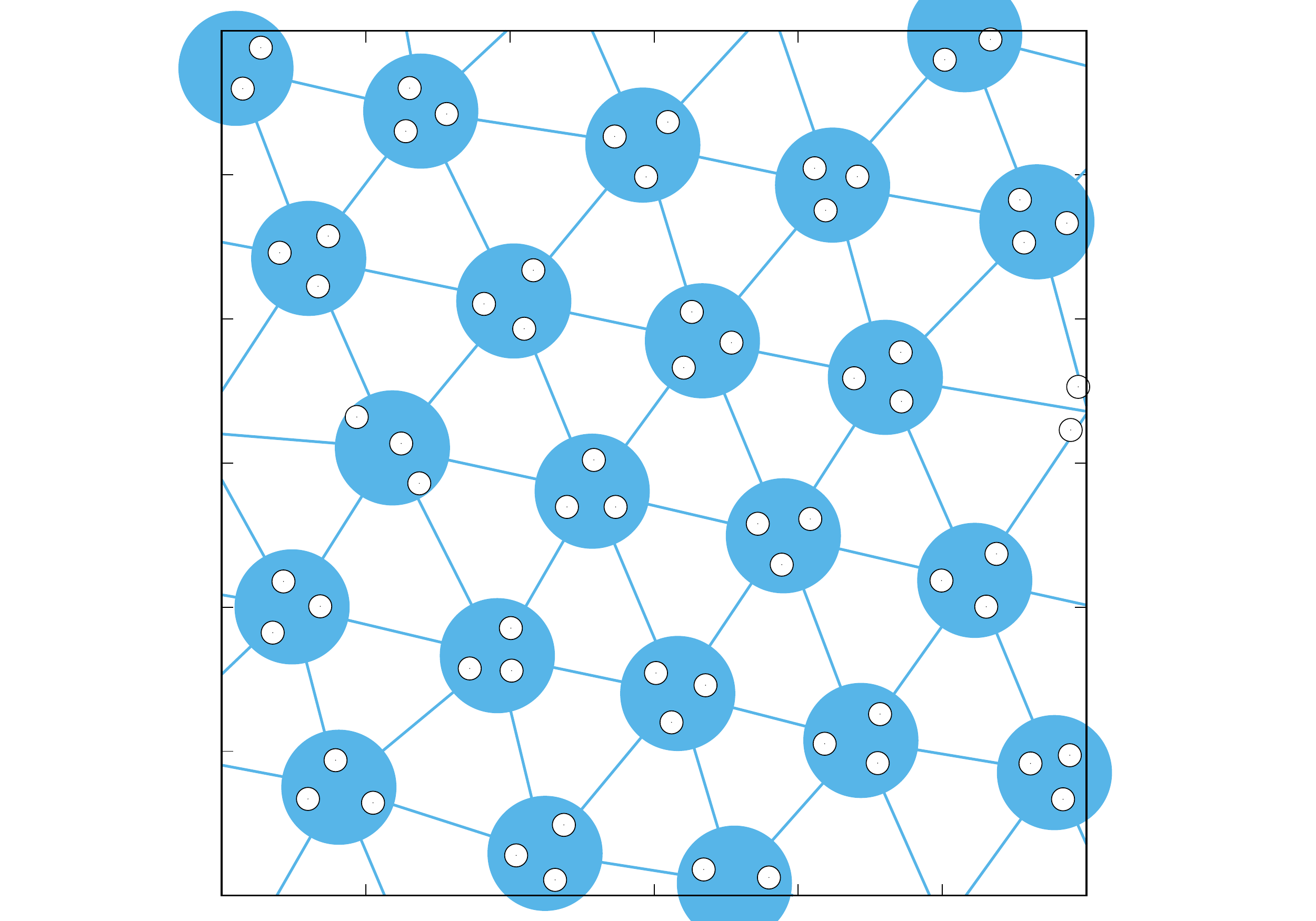}
\caption{}\label{fig:fig_a}
\end{subfigure}
\begin{subfigure}[t]{.4\textwidth}
\centering
\includegraphics[width=\linewidth]{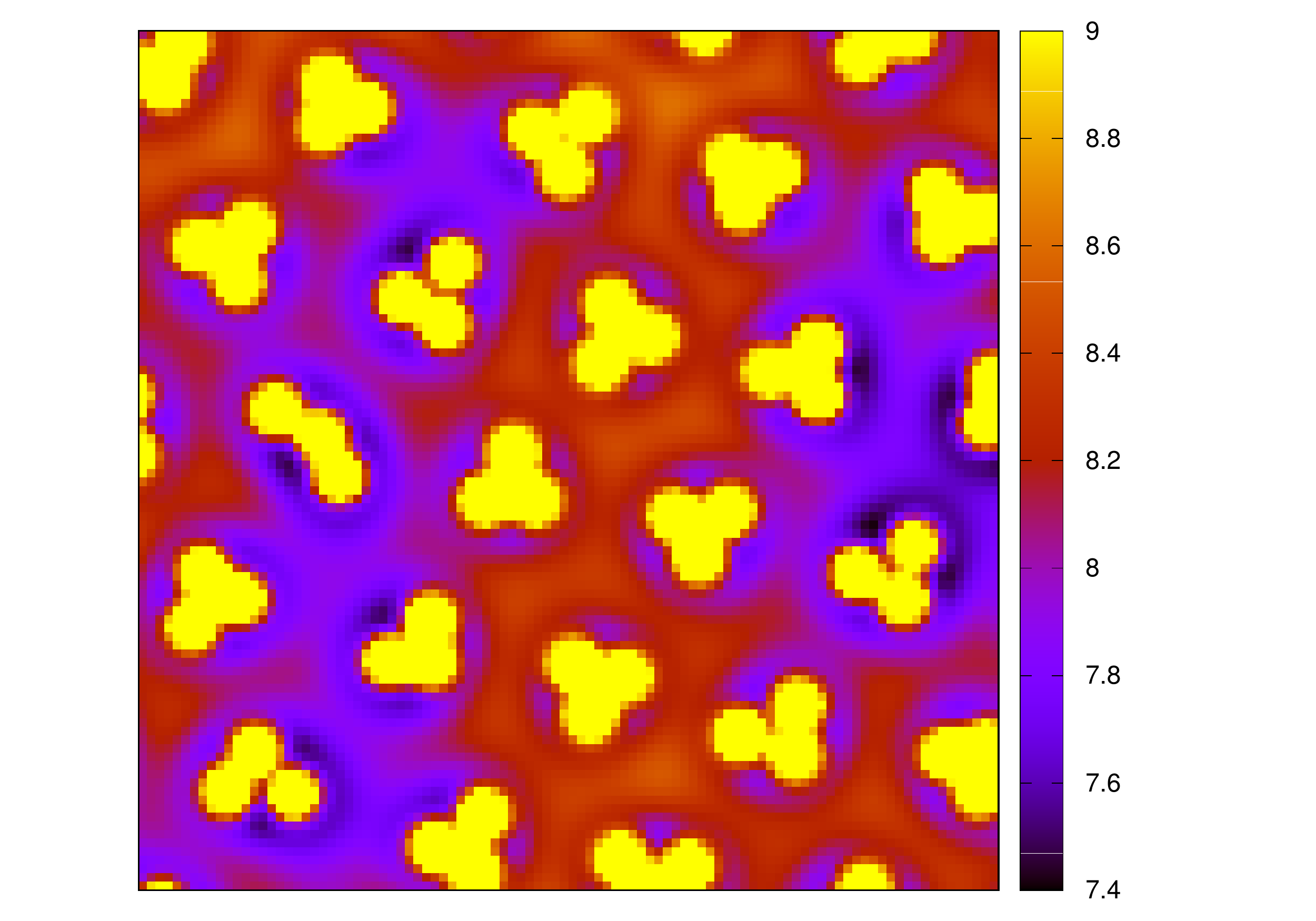}
\caption{}\label{fig:fig_b}
\end{subfigure}
\begin{subfigure}[t]{.4\textwidth}
\centering
\includegraphics[width=\linewidth]{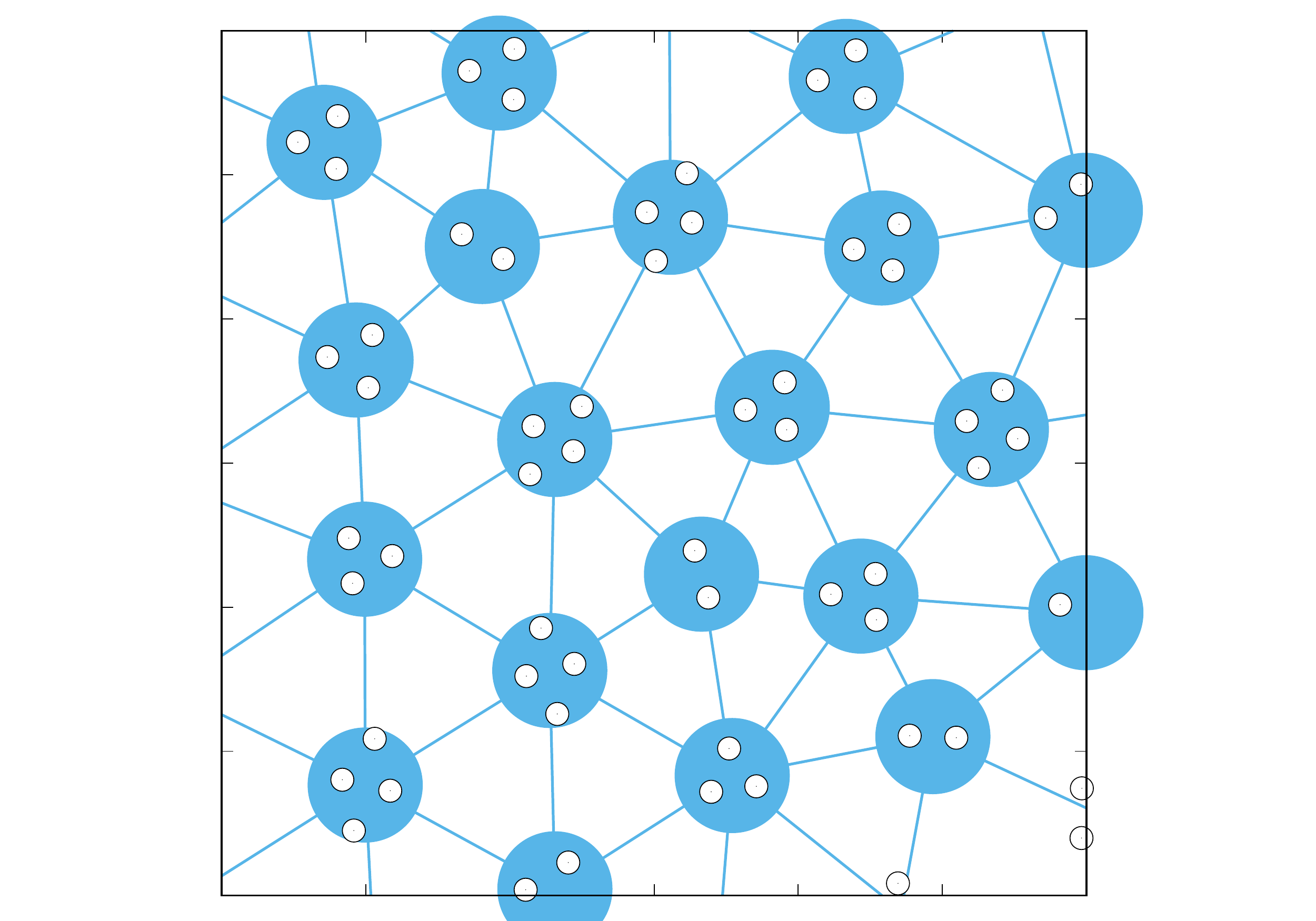}
\caption{}\label{fig:fig_a}
\end{subfigure}
\begin{subfigure}[t]{.4\textwidth}
\centering
\includegraphics[width=\linewidth]{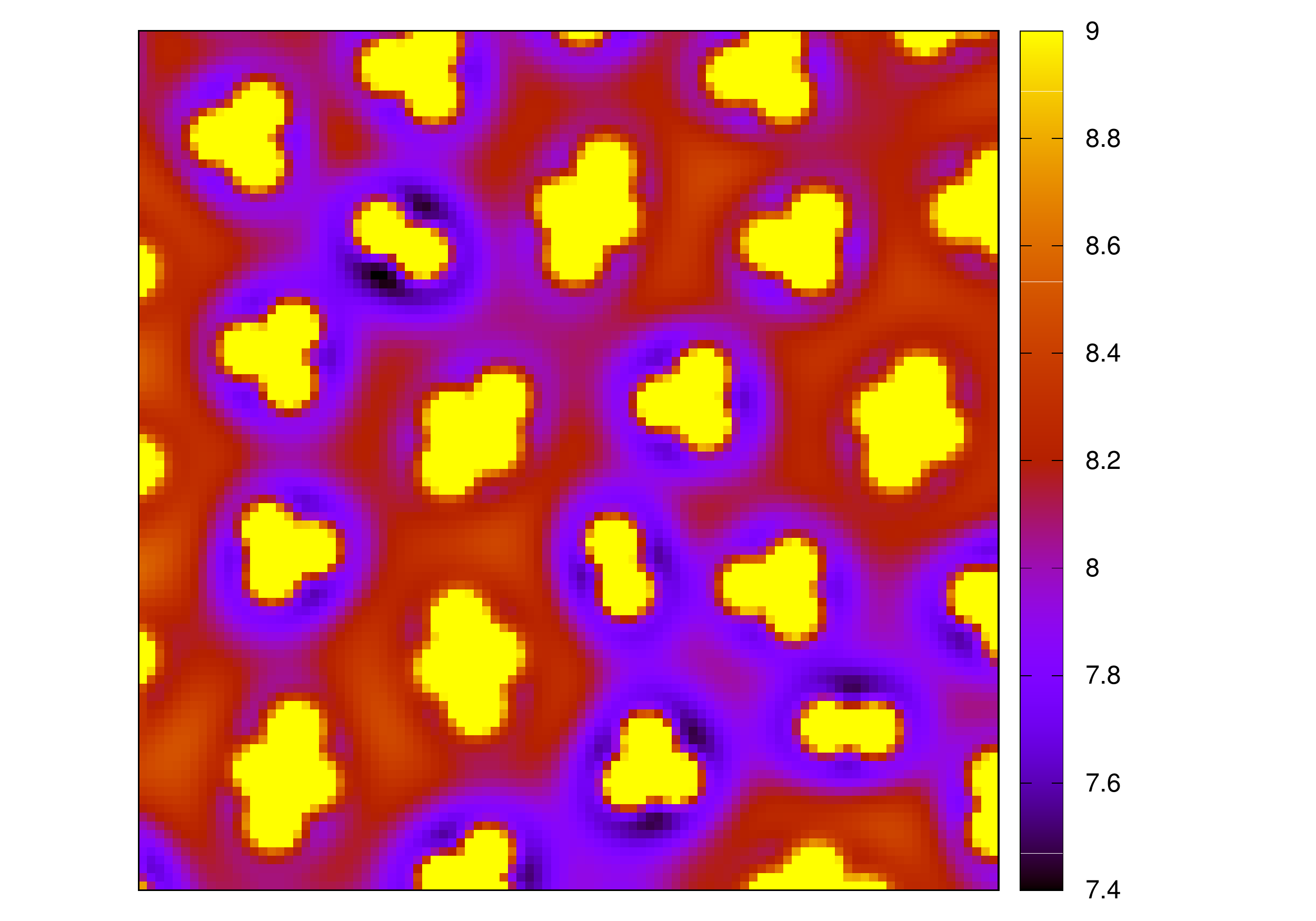}
\caption{}\label{fig:fig_b}
\end{subfigure}
\justify
\caption{Bilayer model in the crystal [a) and b)] and glass [c) and d)] phases. 
a) and c) are snapshots with clusters and nearest-neighbors connections remarked in cyan.
b) and d) are the corresponding maps of potential energy.
}
\label{multi}
\end{figure}

In Fig.~\ref{ultra} and~\ref{multi} we provide the potential energy maps for the ultrasoft and the asymptotic multi-layer vortex potentials obtained by adding a test particle and varying its position, both for the crystal and glass phases, together with snapshots of the corresponding particle configurations for a selected region of space. In all situations, higher occupied sites determine the formation of large energy barriers, which can 
become extended in the case of glasses with large self-induced polydispersity, see, e.g., Fig.~\ref{ultra}. 
This heterogeneous distribution of energy barriers is a general signature for glassy dynamics~\cite{binder2011}. 

The short-range repulsion for the asymptotic vortex potentials determines an additional directionality of the energy barrier for different  particle densities, see Fig.~\ref{multi}. 
As the density increases, this repulsion prevents the system to increase the cluster occupations
[see Fig.~\ref{f4}(c)]. 


\bibliography{c2g}

\end{document}